# Spatiotemporal characteristics of the vertical structure of predictability and information transport in the Northern Hemisphere

A.X. Feng[1, 3], Z.Q. Gong[1, 2], Q.G. Wang[1, 3], G. L. Feng[2, 3, 4]

[1]{College of Atmospheric Sciences of Lanzhou University, Lanzhou, China}

[2]{National Climate Center, China Meteorological Administration, Beijing, China}

[3]{Laboratory for Climate Studies, National Climate Center, China Meteorological Administration, Beijing, China}

[4]{Beijing Normal University, Institute of Global Change and Earth System Sciences, Beijing, China}

Correspondence to: Z. Q. Gong (gzq0929@126.com)

## Abstract

Based on nonlinear prediction and information theory, vertical heterogeneity of predictability and information loss rate in geopotential height filed was obtained over the Northern Hemisphere. In seasonal to interannual time scales, the predictability is low in lower troposphere and high in mid-upper troposphere. But within mid-upper troposphere over the subtropics in some ocean area there is a relatively poor predictability. The conclusions fit the seasonal time scale too. When it goes to the interannual time scale, the predictability becomes high in lower troposphere and low in mid-upper troposphere contrary to the formers. And on the whole the interannual trend is more predictable than the seasonal trend. The average information loss rate is low over the mid-east Pacific, west of North America, Atlantic and Eurasia, and the atmosphere over other places have a relatively high information loss rate in all time scales. Two channels were found steadily over the Pacific Ocean and Atlantic Ocean in subtropics. There exist unstable channels as well. The four-season influence on predictability and information communication was studied too. The predictability is low no matter which season data are removed and each season plays an important role in the existence of the channels except the winter. The predictability and teleconnections are paramount issues in atmospheric science, and the teleconnections may be established by communication channels. So this work is interesting since it reveals the vertical structure of predictability distribution, channel locations and the contribution of different time scales to them and their variations with different seasons.

# 1 Introduction

Forecasting is one of the central problems in atmospheric research considering how the future can be predicted. Three general approaches (Lorenz, 1969) were exploited to solve the question. They are dynamical method, empirical method and dynamical-empirical method. Empirical approach is based on the natural occurrence of analogues, i.e. similar weather situation in meteorology. Farmer and Sidorowich (1987) presented such a nonlinear model on observed data to make predication of chaotic time series. Sugihara and May (1990) simplified and improved the method for making short-term predictions of chaotic dynamical system. Tsonis and Elsner (1992) used the model to distinguish the chaos and random fractal sequences. Elsner and Tsonis (1994) studied the climate predictability in 500hPa applying the approach. And all the above works proved that the model is effective to make predictions of time series from artificial and experimental systems to natural world.

Teleconnections have attracted a lot of interests from scholars and researchers in atmospheric sciences, including early observational studies (Wallace and Gutzler, 1981; Horel and Wallace, 1981), further theoretical/modeling studies [Hoskins and Karoly, 1981; Webster 1981; Webster and Chang, 1988]. Their works revealed the interaction between the tropics and high latitudes and the mechanisms through which the teleconnections might be established. Factors leading to teleconnections are researched as well, such as El Nino. Applying new tool called complex networks to study the atmospheric phenomenon—El Nino revealed that the communication between the tropical and extratropical is less and unstable during the El Nino years (Tsonis and Swanson, 2008; Yamasaki, Gozolchiani and Havlin, 2008). By using the complex networks, the different dynamical mechanisms in the tropics and high latitudes were found (Feng and Gong et al., 2008). So indeed there exists interaction between tropics and high latitudes. The remainings are how to depict aspects of the communication during the interaction and provide the exact locations, where physical entities are exchanged between them. Tsonis and Elsner (1996) had made a great step on the problems by researched the communication transmission at 500hPa, but more work should be done to understand the vertical condition and influences of different time scales and four seasons.

We will focus on describing the heterogeneity of predictability between different pressure levels using the nonlinear method introduced above, and depicting the communication condition by the rate of information loss based on information theory, in order to identify channels in different pressure levels during the interaction between tropics and extratropical. Different time scales and four seasons' influence on the predictability and channels were also studied. The paper will be organized as follow. The data and method are introduced in the section 2. In section 3, vertical distribution of predictability and information communication, different time scale and four season influences on both will be shown. The conclusion is given in the last section.

## 2  Data and method

The data used are from the NCEP/NCAR reanalysis monthly-averaged geopotential height. The pressure levels of 100hPa, 150hPa, 200hPa, 250hPa, 300hPa, 400hPa, 500hPa, 600hPa, 700hPa, 850hPa, 925hPa and 1000hPa are chosen to be study. In each level data cover 1080 grid points in the Northern Hemisphere between 20°N and the North Pole with a latitude-longitude resolution of 5×5 degrees. And the data period is from January 1966 to December 2008. The seasonal and interannual signal is obtained by filtering method proposed by Zheng and Dong (1986). For the study of four-season influences, the data of spring (MAM), summer (JJA), autumn (SON) and winter (DJF) were removed respectively to get the corresponding season's effect for enough data size similarly as Zhi and Gong (2008) did.

The nonlinear prediction model was applied to the above processed data. In each level for a grid point $i$, the first 480 (360 for four-season influence tested) values were chosen as the training set, the remaining values as the target set. As for the training set, an appropriate embedding dimension ($n$) and time delay ($\tau$) were chosen, and the corresponding points in phase space were produced as follows, and vector of the time series is

$$X_i(t) = \{x_i(t),\ x_i(t+t),\ \mathbf{K},\ x_i(t+(n-1)t)\}\ t=1,2,\ldots,516-(n-1)t, i=1,2,\ldots,1080. \quad (1)$$

Here embedding dimension is $n=5$ and time delay is $\tau=2$ as the same as Tsonis and Elsner (1996). The target set is used for prediction following Tsonis and Elsner (1996). In each prediction, the maximum leading time is 12 steps ( i.e. a year). For the same leading time, the correlation coefficient were calculated between predicted and observed values with Pearson correlation coefficient, and the average of all the coefficients was taken in order to view as the grid point's predictability. Thus the 1080 grid points' predictability in different pressure levels is obtained.

Wales (1991) demonstrated that the Kolmogorov entropy ($K$) was related to the initial decay of Pearson correlation coefficient ($r$) and the relationship first presented by Tsonis and Elsner (1996) between them is

$$r(t) = 1 - Ce^{2Kt}. \qquad (2)$$

Based on information theory, the first derivative of information content (i.e. the mean loss of information per unit time) is equal to the negative of Kolmogorov entropy, namely $K \equiv \lim_{T \to 0} \lim_{l \to 0} \lim_{m \to \infty} \frac{1}{mT} \sum_{j=0}^{m-1}(k_{j+1} - k_j) = -\lim_{T \to 0} \lim_{l \to 0} \lim_{m \to \infty} \frac{1}{mT} \sum_{t_0 \cdots t_m} p_{t_0 \cdots t_m} \ln p_{t_0 \cdots t_m}$ where $p_{t_0 \cdots t_m}$ is the joint probability, when phase space divided into D-dimensional hypercubes of content $l^D$ and a trajectory in hypercube $i_0$ at $t=0$, $i_1$ at $t=T$, $i_2$ at $t=2T$, etc. So

$$r(t) = 1 - Ce^{2Kt} = 1 - Ce^{-2\dot{K}t}, \qquad (3)$$

where $C$ is a positive constant. For a period time $t$ the mean value of Pearson's correlation coefficient is

$$<r> = \frac{1}{t}\int_t (1 - Ce^{-2\dot{K}t})dt. \qquad (4)$$

The angular brackets represent taking average of $r$. So the mean coefficient is related to the information loss rate via a function of $f$ (i.e. $<r> = f(\dot{K})$). For a given longitude, the derivative of Eq. (4) with respect to latitude was

$$\frac{d<r>}{dl} = \frac{df(\dot{K})}{dl} = \frac{df(\dot{K})}{d\dot{K}} \frac{d\dot{K}}{dl}. \qquad (5)$$

It is shown that the north-south gradient of $<r>$ is proportional to the north-south gradient of information loss rate. Here the information of those atmospheric phenomena can be exchanged between the tropics and mid-high latitudes, such as El Nino, the 40-50 day oscillation, and so on. If the north-south gradient of <r> is small, it means the rate of information loss along the latitude is slow and the information can flow easily along the longitude. Otherwise, the rate of information loss is fast and the information can not transport easily between the tropics and mid-high latitudes.

## 3 Results

Using the above data and method, we will study the predictability and communication of the atmosphere, and their variations induced by different time scales and four seasons in the Northern Hemisphere.

### 3.1 Vertical characteristics of predictability and information communication

The vertical profile of seasonal to interannual predictability is shown in Fig.1. Setting 30°N be the boundary between the tropics and mid-high latitudes, then a profile is also drawn (Fig.2). All results have been smoothed using a nine-point uniform filter. The Pearson correlation coefficient is the linear correlation between observed and predicted values ranged from 0 to 1. If the correlation is close to 1, it means these areas are more predictable, otherwise the place is less predictable. So from Fig.1 we have found out that on the whole the predictability grows with altitude increase, that is to say, the lower pressure levels (1000hPa to 700hPa) have a poor predictability with Pearson correlation coefficient ranged from 0.4 to 0.7, and the high pressure levels (600hPa-100hPa) have a better predictability ranged from 0.7 to 0.8. The finding fact is easy to be explained, because the lower levels are affected by rather complex surface and strong convective making the cycle signal not obviously. Therefore they are harder to predict and the outcomes are not desirable. While the high levels are relative stable and predicted easily. The Fig.2 is similar to the Fig.1 except for four regions marked by rectangles with black lines. In rectangle 1 and 2 there are poorer predictability compared with other regions along the same altitude. The two regions are just over the Pacific Ocean, and the Gulf of Mexico and Atlantic with longitude from about 120°W to 150°E and 20°W to 80°W respectively. It can be contributed to the strength of the annual cycle just as Elsner and Tsonis (1994) and Tsonis and Elsner (1996) suggested. The atmosphere over large ocean regions is not sensitive to the seasonal cycle, because sensible heat from water can modulate tropospheric temperature and lead to it rather steady. The hot Gulf stream can also make the tropospheric temperature in the region 2 steady, so the region is not sensitive to the seasonal cycle as well. Meanwhile the high levels in troposphere are less affected by surface, so the poorer predicatability are obviously observed in the regions over the ocean compared with those over continent. As for the area of rectangle 3, it has a high predictability than other regions along same altitude. The outcomes may be not credible, because a lot of geopotential heights in 1000hPa are minus because of high terrain of Tibet Plateau (74°E to 104°E, 25°N to 40°N ) located in the region. The area of rectangle 4 is just above Tibet Plateau, and has a poorer prediction may be contributed to the complex surface of the plateau.

In Fig.3 we displays the average rate of the information loss per distance along longitudes according to the Eq. (5), which $\frac{df(\vec{R})}{dl}$ is proportion to $\frac{d<r>}{dl}$ (for $\frac{df(\vec{R})}{d\vec{R}}>0$). If $\frac{d<r>}{dl}$ is close to zero, then the information is not easy to be lost. Therefore the information can easily flow along longitudes from tropics to mid-high latitudes or the opposite direction. If the absolute value of $\frac{d<r>}{dl}$ is much greater than zero, then the information loses quickly. So the information doesn't flow easily along longitudes. The areas covered by two light blue colors (with low information loss rate) are the regions, where the information can be easily transported between tropics and extratropical. A narrow one locates in the mid-upper troposphere (from about 600hPa to 100hPa) over the mid-east Pacific and west of North America (about -160° to -100°, i.e. 100°W to 160°W). Another wide region is over Atlantic and Eurasia with different altitude range under different longitude range (from about 60°W to 60°E the altitude range from about 500hPa to 100hPa, while from about 60°E to 120°E the altitude range from about 300hPa to 100hPa). The remaining areas covered by light purple to dark purple and light pink to dark red are the regions where information can not be easily transported. Tsonis and Elsner (1996) pointed out two communication corridors laid over the Pacific Ocean and Atlantic Ocean at 500hPa in the subtropics. So the profile of information loss rate in subtropics is also drawn to identify channels, and the result is illustrated by Fig.4. The narrow channel over the Pacific exists from about 500hPa to 100hPa but the center moves from about 160°W to 130°W. The center of the wider area over Atlantic moves from about 20°W to 15°W. A new narrow channel was found from about 500hPa to 100hPa centered 60°E. The channels and their location are of importance because they provide the exactly locations in different levels, where information exchanges between the low and the mid-high latitudes and through which they interact each other to affect the global weather and climate.

## 3.2 Predictability and channels in seasonal and interannual time scales

In order to know the predictability of different time scales and their role in tropics and extratropical exchange, the original data was filtered into seasonal and interannual components with the multi-stage filter (Zheng and Dong, 1986). The two components' predictabilities are shown in Fig.5 (a) and (b) respectively. Both of their predictability have a strong relationship with altitude, but the seasonal predictability grows with increasing altitude but the interannual one is just opposite. Meanwhile the interannual predictability is better (ranged from 0.89 to 0.92) than the seasonal one (from 0.4 to 0.8). That means the interannual trend is more predictable than the seasonal trend. The pattern of seasonal predictability is similar to that of the seasonal to interannual, indicating the seasonal cycle plays a great role in climate system.

Fig.6 displays the number and the location of channels in different levels of the subtropics in the two time scale. Both of the two time scale contribute a lot to the tropics and high latidudes exchange, because the two channels over Pacific Ocean and Atlantic Ocean steadily exist in the two time scale except a change in altitude range and width (comparison can be seen from Table.1). So the exchange between tropical and extratropical happens from seasonal to interannual time scale, from our study through the two channels over ocean that the teleconnections influencing the global weather and climate are established. There exist unstable channels too, such as the one over Tibet Plateau from 400hPa to 100hPa with longitudes from about 80°E to 120°E, existing only in interannual time scale, and the narrow channel centered 60°E, which may come into being through interaction between seasonal and interanuual for it existing only in seasonal to interanuual time scale. Meanwhile in seasonal time scale the information mainly transports from tropics to high latitudes (with positive information loss rate) but the interannual are opposite mainly from high latitudes to tropics (i.e. with negative information loss rate), which can be obviously seen from the Fig.6. This is very interesting.

## 3.3 Four-season influences on the predictability and communication channels

The predictability and information communication in the Northern Hemisphere are closely related to seasonal cycle. In order to know the role of different seasons in predictability and information communication, spring, summer, autumn and winter data were artificially removed respectively. For example, to study the contribution of spring, geopotential heights of March, April, and May were removed from the time series from 1966 to 2008. Then the above theory and procedure were applied to the remaining data.

Fig.7 shows the predictability when spring, summer, autumn, winter data were removed respectively. Fig.8 shows the difference of prediction between the whole and the remaining data. The predictability would be low when any season data were removed, especially for the mid-upper troposphere, which indicates that the annual circle is a standard circle of a major external forcing. Among the seasons, spring has the strong influence on the predictability, followed by the autumn and summer, and the winter's influence is the smallest. For spring and autumn, the reduction of predictability is somewhat related to the altitude. The influence of summer and winter seems to different effect to different longitude range.

The seasons have great influence on the channels in the subtropics except winter (Fig.9). Spring has the greatest impact on the channels, almost all three channels disappear when the data of it removed. The influence of summer mainly centers on the disappearance of the channel over the Atlantic Ocean and the one centering 60°E, and is less effective to the channel over Pacific Ocean. The narrow channel centering 60°E is not sensitive to autumn, but the other two are almost disappearance when the autumn data removed. All the three channels are less effect by winter compared with other seasons just a change in altitude range and width compared with the original ones. In other word spring contribute more to the north-south information exchange through controlling the corridors' opening in subtropics, summer controlling the Atlantic Ocean channel and the one centering 60°E as well, and autumn controlling the Pacific Ocean one and the Atlantic Ocean one. The winter has a little impact on the channels that maybe why all the teleconnection patterns steadily existing in winter. So it is of great importance to realize the role of seasons in meridional information communication and teleconnections. The channels affect by different time scales and seasons can be obtained in Table 1.

## 4  Conclusions

Study of the predictability and the information communication between 20°N and the North Pole in troposphere on the basis of nonlinear prediction and information theory has been made. The predictability is low in lower troposphere and high in mid-upper troposphere in seasonal and seasonal to interannual time scales. Oppositely, the interannual predictability is high in lower troposphere and low prediction in mid-upper troposphere. The predictability of interannual time scale is better than that of seasonal indicating the interannual trend is more predictable than the seasonal trend. That fits our common sense. Meanwhile the predictability is low no matter what season data removed, what verifies the annual cycle play a great role in large scale dynamics. Among the four-seasons the impact of winter is the smallest and spring is the biggest to the predictability of the atmosphere.

The study of the vertical information communication indicated that the average information loss rate is low in the mid-upper troposphere over mid-east Pacific, west of North America, Atlantic and Eurasia, and high in atmosphere over mid-west Pacific, east of America and low troposphere over Asia. Two channels were found steadily over the Pacific Ocean and Atlantic Ocean in subtropics, which are contributed to both the seasonal and interannual time scale signal. The seasonal to interannual unstable channel centered 60°E in upper troposphere may come into being through interaction between seasonal and interanuual signals. The interannual unstable channel locates from 400hPa to 100hPa with longitude from about 80°E to 120°E. The seasons play an important role except winter in impacting on the existence of the channels, which controlling the information exchange between the tropics and the mid-latitudes, especially the spring. The less influence on the channels in winter may explain why patterns of teleconnection steadily exist in winter.

One important thing is that this study demonstrates the Tibet Plateau plays a great role in predictability and information communication in the Northern Hemisphere. And the most interesting thing is that in seasonal time scale the information mainly transports from tropics to high latitudes but the interannual is just opposite.

## Acknowledgements

The authors thank anonymous reviewers and editors for beneficial and helpful suggestions for this manuscript. We would also like to thank Prof. Anastasios Tsonis for providing materials on applying nonlinear predication method. This work is jointly supported by the National Natural Science Foundation of China (40930952, 40875040, 40775048), the Special project for Public Welfare Enterprises (GYHY200806005) and the National Science/Technology Support Program of China (2007BAC29B01).

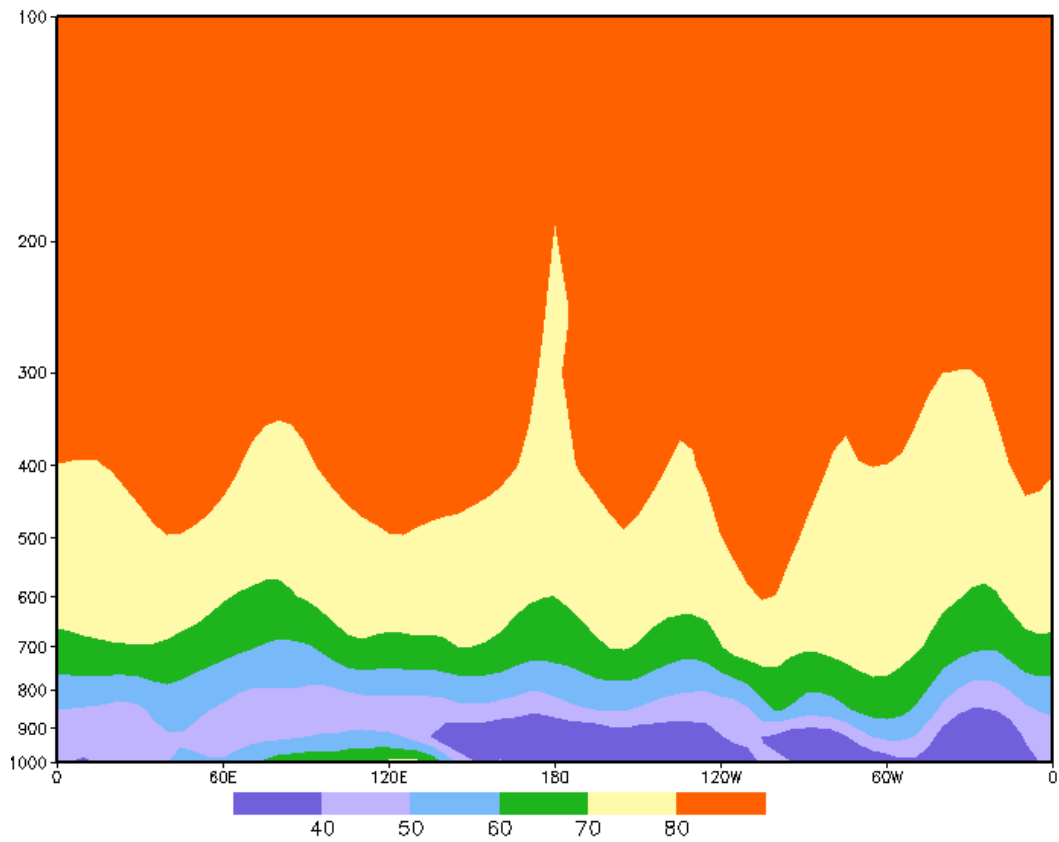

Figure 1. Pearson correlation coefficient (×100) distribution from 1000hPa to 100hPa, which was averaged between 20°N and the North Pole for each pressure level.

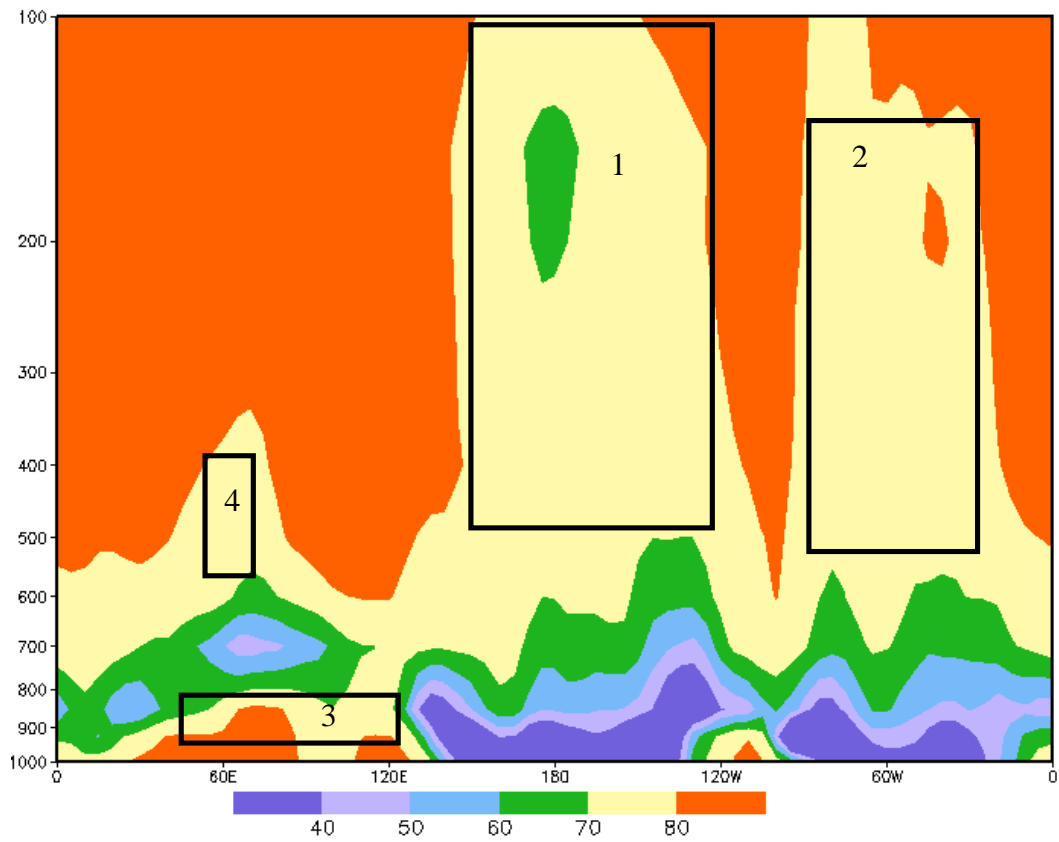

Figure 2. Pearson correlation coefficient (×100) distribution along 30°N from 1000hPa to 100hPa.

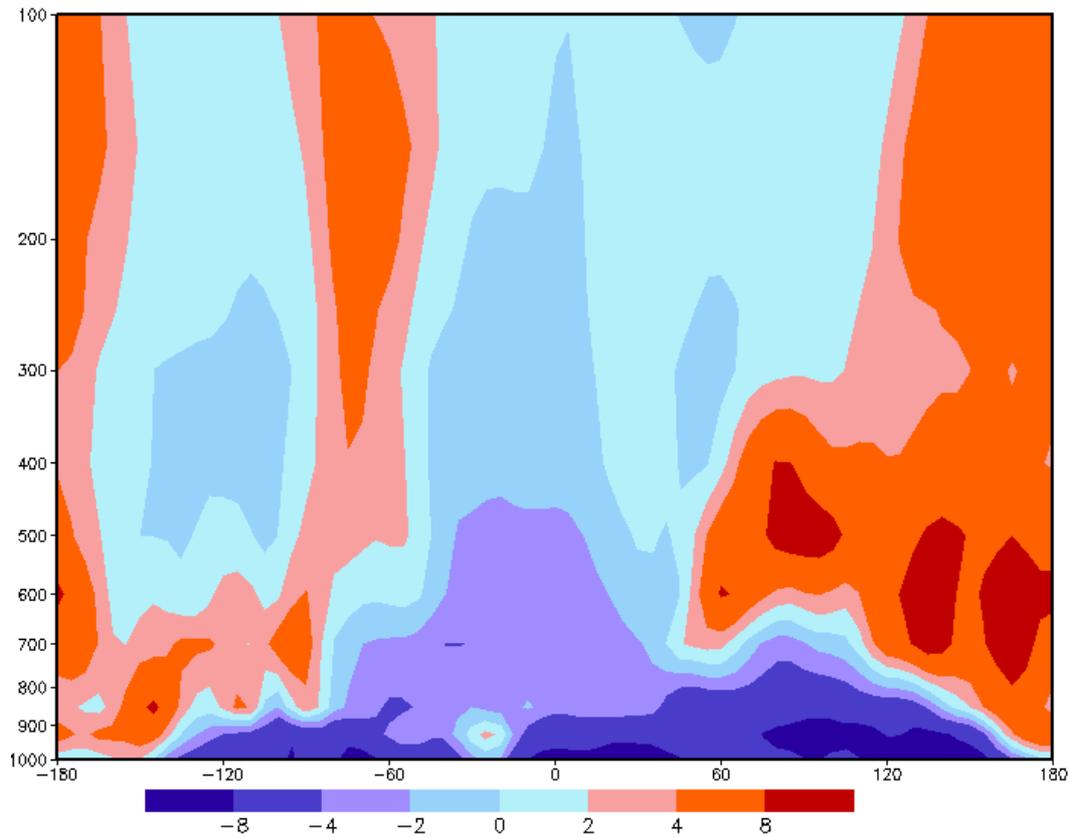

Figure 3. Taken north-south gradient of Pearson correlation coefficient (×100) in each level and done the same latitude average as Fig.1 to obtain the figure. It indirectly displays the average rate of the information loss per distance along longitudes. The regions shaded by two light blue colors are the locations where information flows easily, and the other regions are the locations information not flowing easy. In order to put the regions of low information loss rate together, the abscissa is not the same as Fig.1 and Fig.2, negatives represent west longitudes and positives are east longitudes. And the gradient greater-than-zero means the information flowing from tropics to high latitudes, vice versa.

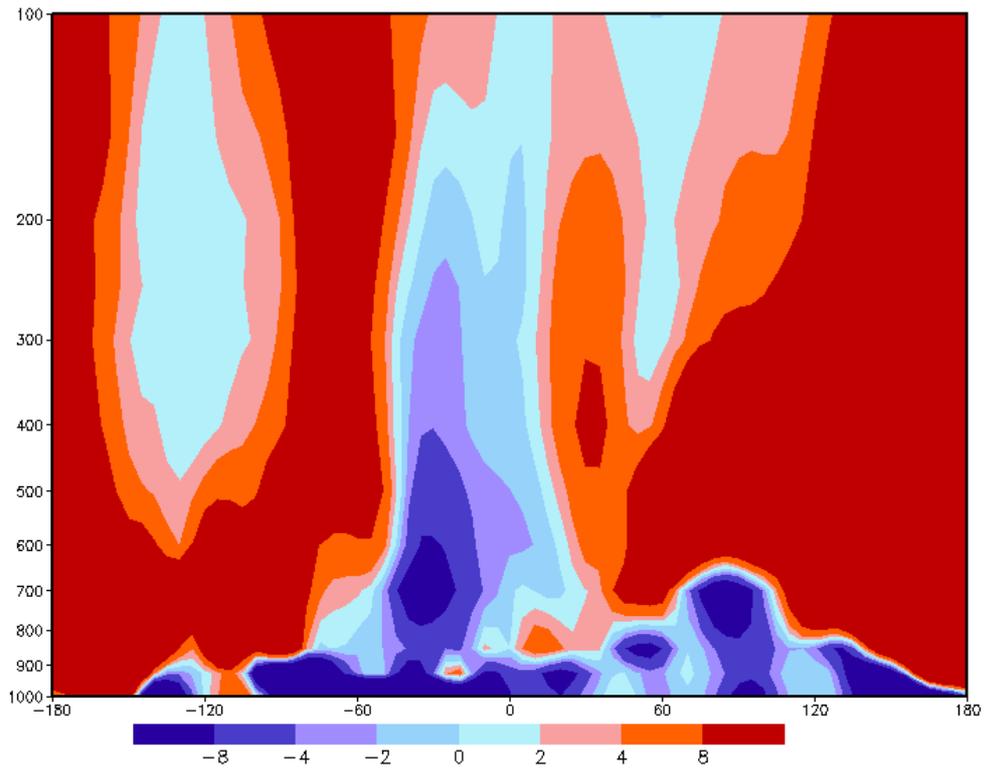

Figure.4 The same as Figure.3 but averaged between 20°N and 30°N , the regions covered by light blue colors are the channels where most information exchanges between tropics and extratropical in different levels.

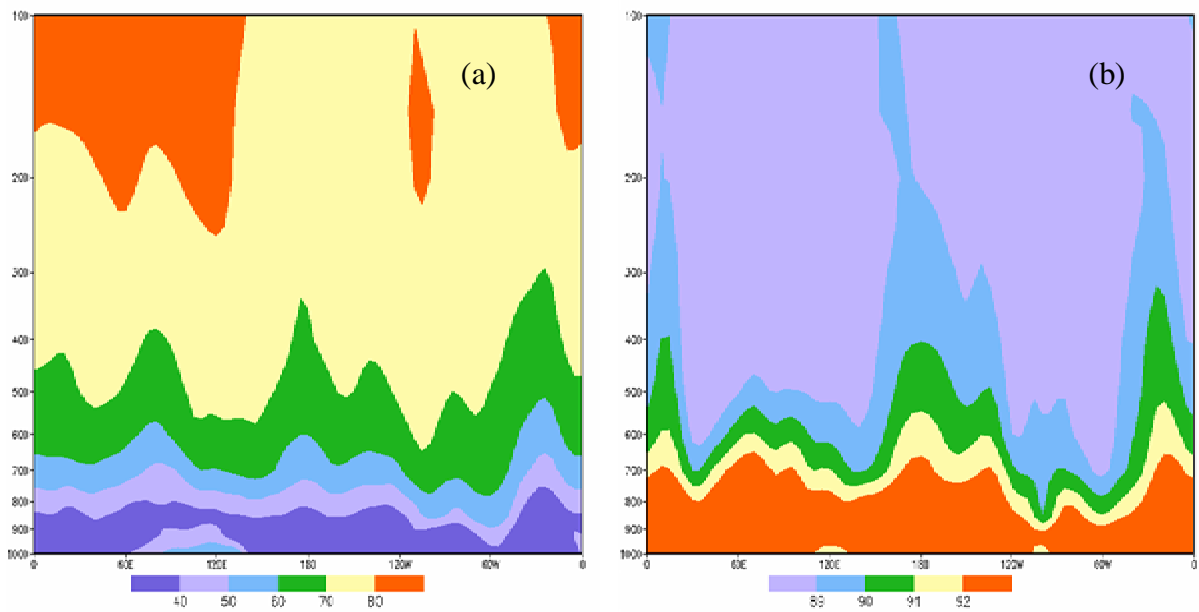

Figure.5 Just like Figure.1 but in seasonal (a) and interannual (b) time scale: the predictability of interannual signal is better than that of seasonal signal, and predictability of the former decreases with altidude increasing but the latter is oppositely.

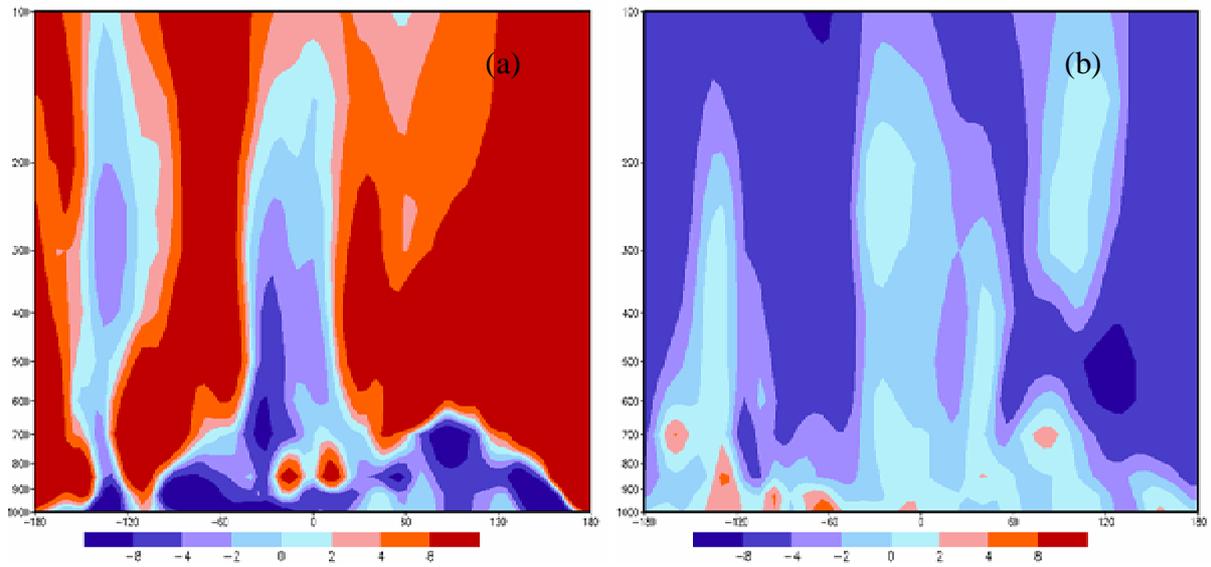

Figure.6 The same as Figure.4 identifying the channels in seasonal (a) and interannual (b) time scale along longitudes in subtropical: both of the two time scale contribute a lot to the two channels over Pacific Ocean and Atlantic Ocean, while the interannual time scale plus one more channel from 400hPa to 100hPa with longitudes from 80°E to 120°E.

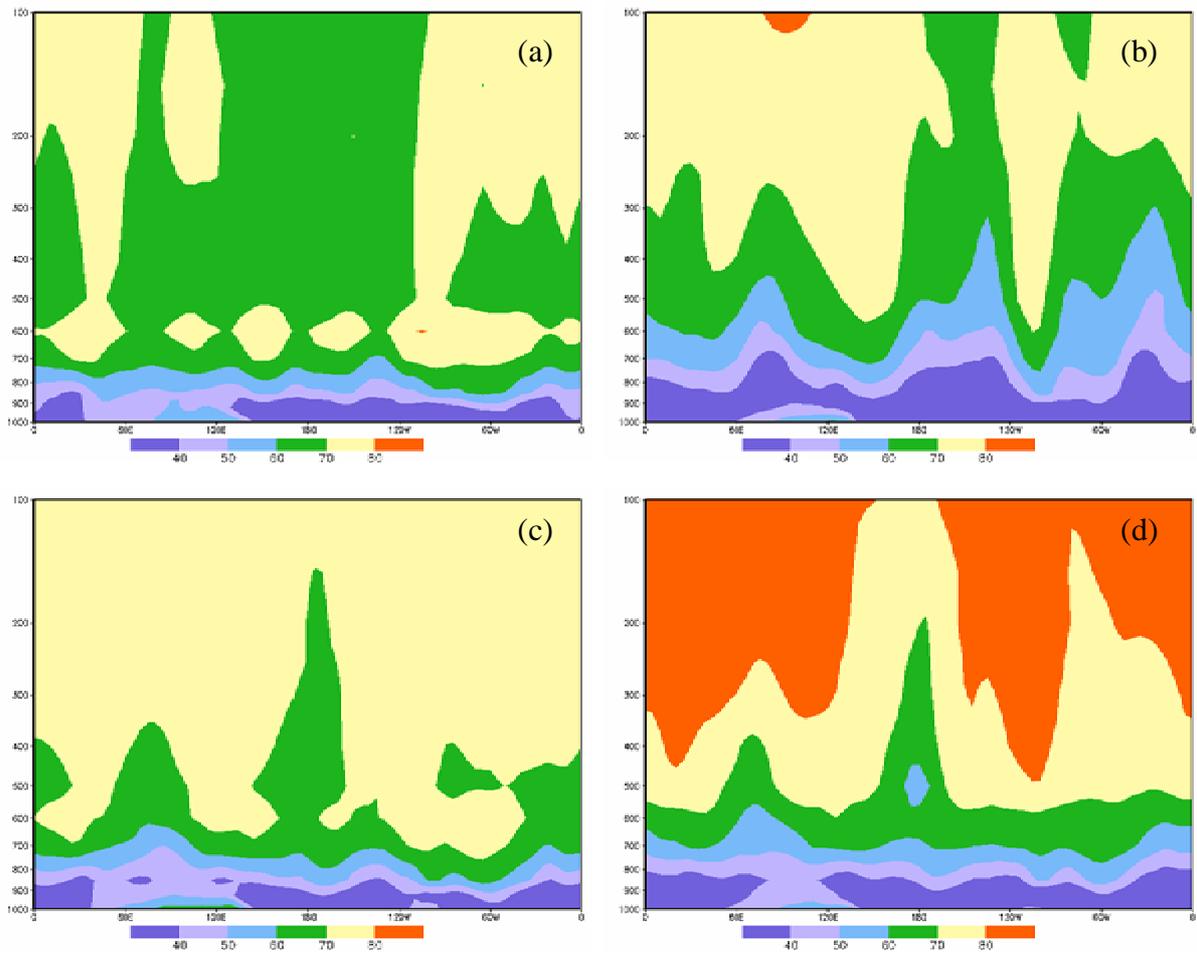

Figure.7 Just as Figure.1 but referring to the situation when (a) spring data removed; (b) summer data removed; (c) autumn data removed; and (d) winter data removed.

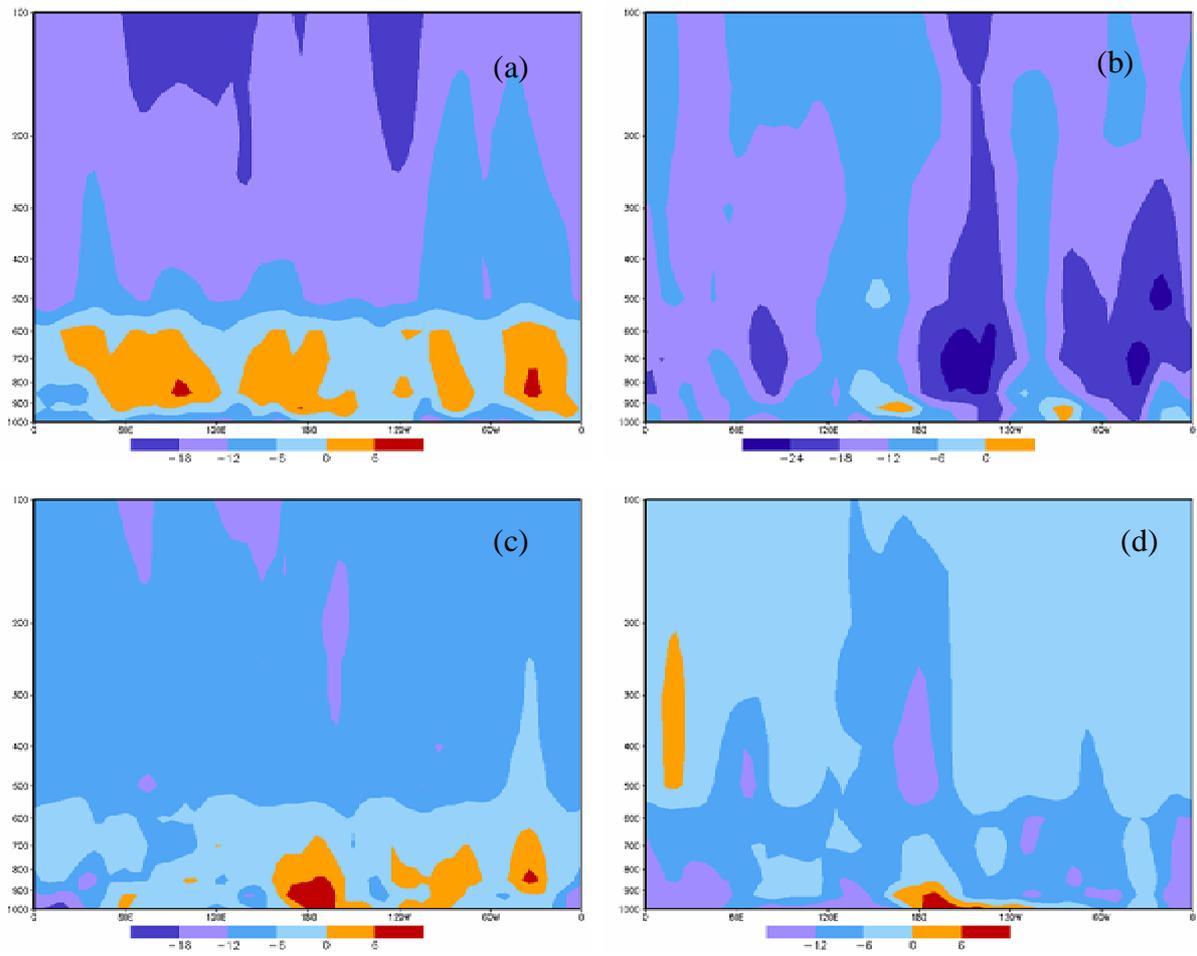

Figure.8 The difference between Figure.1 and Figure.7: (a) Figure.7(a) subtracts Figure.1; (b) Figure.7(b) subtracts Figure.1; (c) Figure.7(c) subtracts Figure.1; and (d) Figure.7(d) subtracts Figure.1.

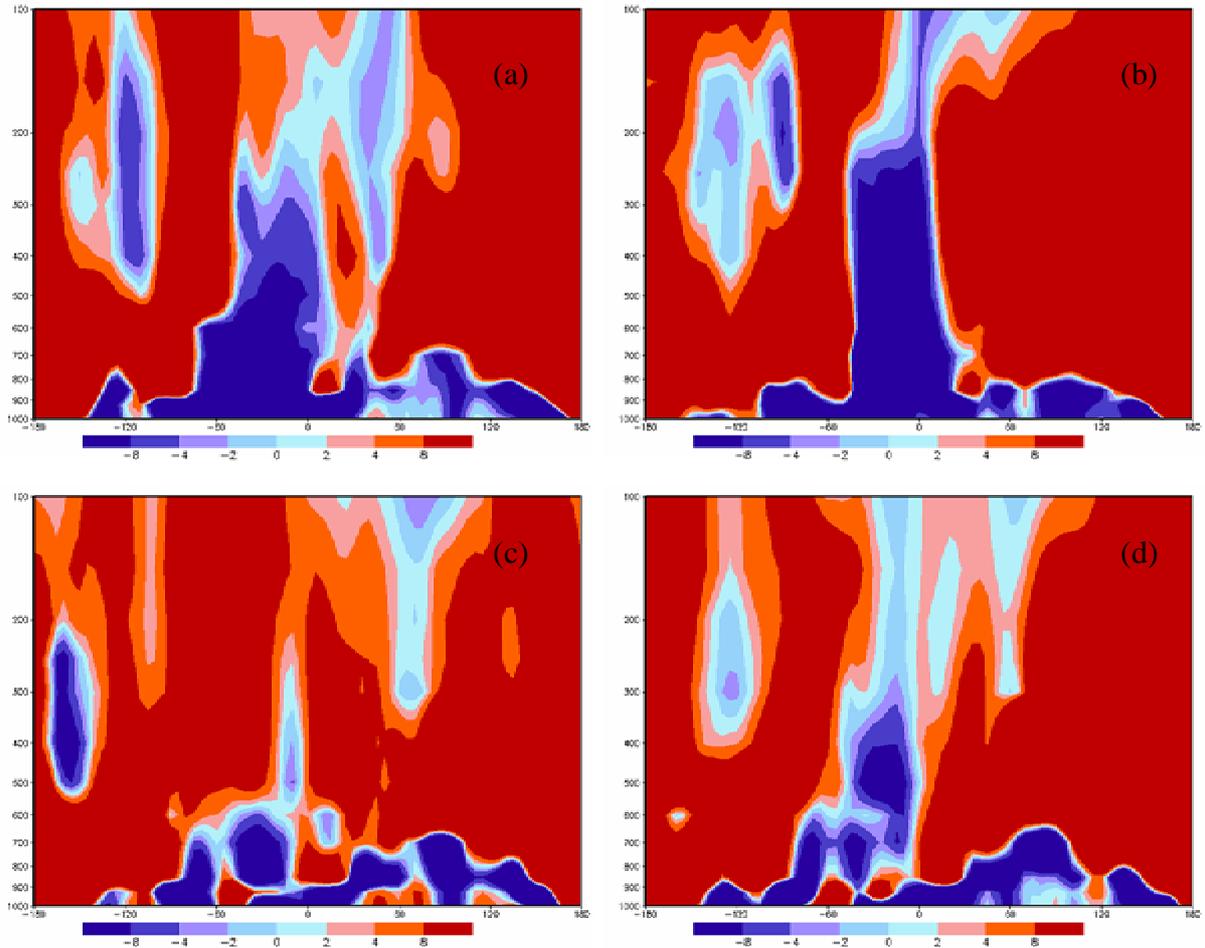

Figure.9 The effect of seasons to channels in subtropics: (a) the three channels are sensitive to the spring, when its data are removed almost all the channels disappearance; (b) only the channel over Pacific Ocean is not very sensitive to summer; (c) the narrow channel centered 60°E is not very sensitive to autumn but the other two; and (d) none of the three channels is very sensitive to the winter.

**Table 1.** The channels in different time scale, affected by different seasons, and their about location.

|  | number | channel | altitudes | longitudes (center) |
|---|---|---|---|---|
| Seasonal to interannual | 3 | over Pacific Ocean | 500-100hPa | 110°W-150°W(130°W) |
|  |  | over Atlantic Ocean | 500-100hPa | 45°W-15°E(15°W) |
|  |  | the narrow one | 500-100hPa | 55°E-65°E(60°E) |
| Spring data removed | 0 |  |  |  |
| Summer data removed | 1 | over Pacific Ocean | 500-150hPa | 115°W -150°W(130°W) |
| Autumn data removed | 1 | the narrow one | 400-100hPa | 55°E-75°E(65°E) |
| Winter data removed | 3 | over Pacific Ocean | 400-150hPa | 110°W-140°W(125°W) |
|  |  | over Atlantic Ocean | 300-100hPa | 40°W-0(20°W) |
|  |  | the narrow one | 300-100hPa | 55°E-65°E(60°E) |
| Seasonal time scale | 2 | over Pacific Ocean | 700-100hPa | 115°W -150°W(130°W) |
|  |  | over Atlantic Ocean | 400-150hPa | 45°W-10°E(15°W) |
| Interannual time scale | 3 | over Pacific Ocean | 700-200hPa | 120°W -150°W(135°W) |
|  |  | over Atlantic Ocean | 1000-150hPa | 45°W-10°E(15°W) |
|  |  | over Tibet Plateau | 400-100hPa | 80°E-120°E(100°W) |